\documentclass[sigconf]{acmart}

\usepackage{colortbl}   
\usepackage{dsfont}     
\usepackage{multirow}   

\newcommand{\yes}{\checkmark}
\newcommand{\pmark}{{\scriptsize(\checkmark)}}
\newcommand{\artifacturl}{\url{https://github.com/Eureka246/BulkPR-Bench-Release}}

\setcopyright{none}
\acmConference[KDD '27]{ACM SIGKDD Conference on Knowledge Discovery and Data Mining}{August 1--5, 2027}{San Jose, CA, USA}
\acmYear{2027}
\acmMonth{8}
\acmDOI{}
\acmISBN{}
\begin{document}

\title{BulkPR-Bench: Benchmarking Queue-Level Governance of Interacting Pull Requests}

\author{Zetong Xiong}
\authornotemark[2]
\affiliation{%
  \institution{Baidu}
  \city{Beijing}
  \country{China}
}
\email{xiongzetong@baidu.com}

\author{Qiao Zhao}
\authornote{Qiao Zhao, Jun Zhang, Xueying Lyu, Zhi Li, and Yehua Yang
are corresponding authors.}
\authornote{Zetong Xiong and Qiao Zhao contributed equally.}
\affiliation{%
  \institution{Baidu}
  \city{Beijing}
  \country{China}
}
\email{zhaoqiao@baidu.com}

\author{Jun Zhang}
\authornotemark[1]
\author{Xueying Lyu}
\authornotemark[1]
\author{Zhi Li}
\authornotemark[1]
\author{Yixiang Tu}
\affiliation{%
  \institution{Baidu}
  \city{Beijing}
  \country{China}
}
\email{zhangjun25@baidu.com}
\email{lvxueying@baidu.com}
\email{lizhi02@baidu.com}
\email{tuyixiang@baidu.com}

\author{Xiaowen Yang}
\author{Yunjie Zhang}
\author{Yufeng Wang}
\author{Zhe Zhang}
\affiliation{%
  \institution{Baidu}
  \city{Beijing}
  \country{China}
}
\email{yangxiaowen03@baidu.com}
\email{zhangyunjie02@baidu.com}
\email{wangyufeng09@baidu.com}
\email{zhangzhe25@baidu.com}

\author{Kaize Yu}
\author{Hanwen Du}
\author{Zhongkai Sun}
\author{Zhuoxin Liu}
\affiliation{%
  \institution{Baidu}
  \city{Beijing}
  \country{China}
}
\email{yukaize@baidu.com}
\email{duhanwen01@baidu.com}
\email{sunzhongkai@baidu.com}
\email{liuzhuoxin@baidu.com}

\author{Zekun Lin}
\author{Jianwen Yang}
\author{Ruining Chen}
\author{Ying Zhang}
\affiliation{%
  \institution{Baidu}
  \city{Beijing}
  \country{China}
}
\email{linzekun@baidu.com}
\email{yangjianwen02@baidu.com}
\email{chenruining@baidu.com}
\email{zhangying84@baidu.com}

\author{Tingxuan Pan}
\author{Ke Chen}
\author{Shubin Han}
\author{Chuanhao Sun}
\affiliation{%
  \institution{Baidu}
  \city{Beijing}
  \country{China}
}
\email{pantingxuan@baidu.com}
\email{chenke21@baidu.com}
\email{hanshubin@baidu.com}
\email{sunchuanhao01@baidu.com}

\author{Yehua Yang}
\authornotemark[1]
\affiliation{%
  \institution{Baidu}
  \city{Beijing}
  \country{China}
}
\email{yangyehua@baidu.com}

\renewcommand{\shortauthors}{Xiong et al.}

\begin{abstract}
Coding-agent benchmarks increasingly cover long-horizon, end-to-end, and
interactive development, but typically retain one requested outcome or a
fixed change sequence. Sequential policies can process a pull-request (PR)
queue one candidate at a time, but when queued PRs interact, maximizing safe
delivery can require jointly deciding which changes to merge and in what
order.
We introduce
\textbf{BulkPR-Bench}, an executable benchmark in which an agent must recover
consequential PR relations and return a large safe subset in executable order
under a rolling-release protocol. The suite contains 581 newly authored candidate PRs
on frozen snapshots of 18 real repositories.
Registered state-by-state repository execution, including hidden safety
checks, validates the gold relation graph; an exact oracle then computes the
largest safe subset. Our primary metric, \emph{Relational Delivery Score} (RDS), scores
safe delivery and correct rejection over relation groups from the realized
merge trace; Global Safety-Gated Yield (Global-SGY) separately
measures strict delivery of the realized whole-queue plan. Under the buffered primary protocol with
batch size $K{=}32$, the three highest RDS estimates among the six models are
66.6\%, 62.0\%, and 57.9\%, compared with 53.1\% for the strongest sequential
baseline. Only 8 of 324 model runs complete a queue exactly. Critical-relation
recall ranges from 35.2\% to 57.7\%, and diagnostic runs supplied with the
gold relations show substantial remaining headroom. Gains on relation groups
therefore do not yet translate into dependable whole-queue governance.
\end{abstract}



\maketitle

\section{Introduction}\label{sec:intro}

\begin{figure*}[t]
  \centering
  \includegraphics[width=\textwidth]{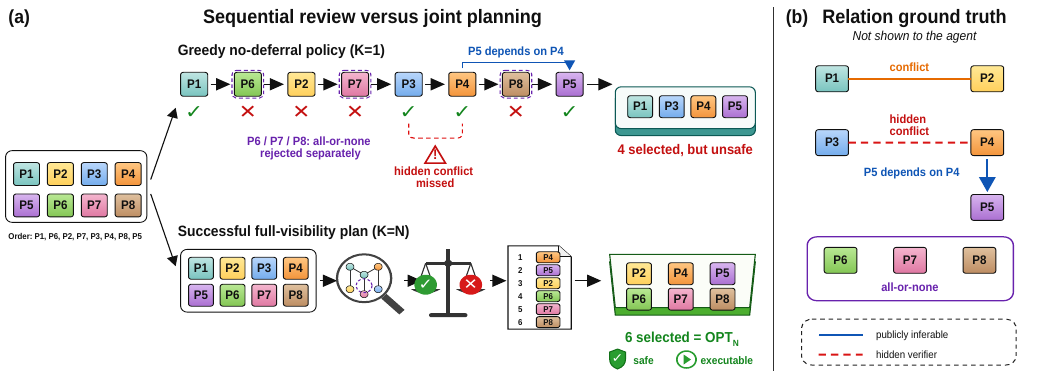}
  \Description{Two-panel illustration of an eight-PR queue. Panel (a)
  contrasts a greedy no-deferral policy at $K{=}1$ with a successful
  full-visibility plan at $K{=}N$. The greedy policy accepts P1, P3, P4,
  and P5 but is unsafe because P3 conflicts with P4; the successful plan
  outputs the safe executable order P4, P5, P2, P6, P7, P8, thereby
  respecting P5's dependency on P4. Panel (b) shows the hidden gold
  relations: P1 conflicts with P2, P3 conflicts with P4, P5 depends on P4,
  and P6, P7, and P8 form an all-or-none group.}
  \caption{In this episode, a greedy no-deferral policy admits a
  hidden-conflicting pair and rejects the all-or-none group; a successful
  full-visibility plan selects the maximum safe six-PR subset in executable
  order. Gold relations remain hidden from the agent.}
  \label{fig:task-overview}
\end{figure*}

Code review and pull-based development normally assess one proposed change
against the current trunk~\cite{bacchelli2013expectations,
rigby2013convergent,gousios2014pullbased,gousios2015integrators,
tsay2014social}. CI studies and production merge queues improve the
reliability or throughput of that process~\cite{hilton2016ci,
vasilescu2015quality,beller2017travis,juloori2025ciatscale}. A queue of
interacting candidates changes the decision, however. Parallel changes can
interfere syntactically or semantically~\cite{perry1998parallel,
ghiotto2020nature,vale2022challenges}; review data also exposes
dependencies, duplicate or competing PRs, and concurrent edits
~\cite{arabat2024dependencies,li2022duplicateprs,zhang2018competingprs,
maddila2022cone}. A sequential public-gate policy can process a queue and
keep public CI green, yet still merge a combination that fails hidden safety
checks, reject changes that are valid only together, or deliver fewer safe
changes than a joint plan.

We study \emph{queue-level PR governance}: recover the relations that matter,
choose which candidates to merge or reject, and order the chosen subset so
that repository execution succeeds. BulkPR-Bench makes this a repeatable
task. Each instance pairs a pinned repository snapshot with $N$
newly authored candidates revealed in batches of size $K$. The agent submits
atomic merge proposals, deferrals, and a revisable relation ledger. Scoring
follows the realized plan accepted by the executor, not the agent's claimed
plan. Hidden verifiers, an execution-validated gold relation graph, and exact
safe-subset optima provide the scoring truth.
Figure~\ref{fig:task-overview} shows why arrival-order decisions and joint
planning can produce different subsets and orders.

The evaluation asks four linked questions. \textbf{RQ1 (Final delivery)}
asks how agents compare with deterministic public-CI queue baselines.
\textbf{RQ2 (Visibility and batch size)} asks how $K$ and deferral change
available information and realized decisions. \textbf{RQ3 (Relation recovery
and action)} asks how completely agents recover consequential relations and
what headroom appears when reliable relation information is supplied.
\textbf{RQ4 (Strict completion and outcomes)} asks how often exact
whole-queue completion occurs and how remaining runs terminate.

This paper contributes four pieces of evidence and infrastructure.
(1)~The \emph{queue-level governance task} couples relation recovery, subset
selection, executable order, and bounded rolling visibility
(Section~\ref{sec:task}).
(2)~\emph{Realized-plan measurement} uses RDS as the sole ranking metric,
with whole-queue, hidden-relation, recovery, and action diagnostics
(Section~\ref{sec:metrics}).
(3)~An \emph{executable relational construction} draws on historical PRs and
expert input to author new candidates on real snapshots, and establishes
their relations and safe optima through repository execution
(Section~\ref{sec:construction}).
(4)~A formal evaluation over 18 repositories finds that the three highest
buffered model RDS estimates are 66.6\%, 62.0\%, and 57.9\%, compared with
53.1\% for the strongest sequential baseline; only 8 of 324 model runs
exactly complete a queue (Section~\ref{sec:results}).

\section{Related Work}\label{sec:related}
Existing work covers individual components of queue-level governance.
Table~\ref{tab:benchmark-comparison} compares representative coding-agent
benchmarks and integration methods by the inputs, outputs, and decisions they
evaluate.

%
%

\begin{table*}[t]
\centering
\normalsize
\renewcommand{\arraystretch}{1.0}
\setlength{\tabcolsep}{3.25pt}
\setlength{\heavyrulewidth}{1pt}
\caption{Comparison of representative coding-agent benchmarks and integration
methods. Marks describe properties of the evaluated task or method, not model
performance; the axes are defined below.}
\label{tab:benchmark-comparison}
\begin{tabular}{@{}lllcccccc@{}}
\toprule
Work & Input / \# changes & Primary output & Multi & Exec & Comp & Subset & Order & Roll-$K$ \\
\midrule
\multicolumn{9}{@{}l}{\textbf{Repository-level \& sequential-evolution anchors}}\\
SWE-bench~\cite{swebench}                           & 1 issue            & 1 patch                &      & \yes &      &      &      &       \\
DeepSWE / SWE-Marathon~\cite{deepswe,swemarathon}   & 1 objective        & working implementation &      & \yes &      &      &      &       \\ 
SWE-Cycle~\cite{swecycle}                           & 1 issue / 3 phases & env. + tests + patch   &      & \yes &      &      &      &       \\ 
Dialogue SWE-Bench~\cite{dialogueswebench}          & 1 issue + dialogue & 1 patch                &      & \yes &      &      &      &       \\ 
SWE-STEPS / ChainSWE~\cite{beyondisolated,chainswe} & fixed change chain & cumulative implementation(s) & \yes & \yes & \yes &      &      & \pmark \\ 
\addlinespace[2pt]
\multicolumn{9}{@{}l}{\textbf{PR / code review}}\\
c-CRAB / SWE-Review~\cite{crabench,swereview}       & 1 PR (+issue)      & review / decision      &      & \yes &      &      &      &       \\ 
\addlinespace[2pt]
\multicolumn{9}{@{}l}{\textbf{Composition \& integration}}\\
MOSAIC-Bench~\cite{mosaic}                          & 3 fixed stages    & cumulative diff         & \yes & \yes & \yes &      &      & \pmark \\ 
MCon4J~\cite{mcon4j}                                & 3-way / octopus merges & generated conflict tests & \yes & \yes & \yes &      &      &       \\
GitGoodBench~\cite{gitgood}                         & branches/commits  & resolved git state      & \yes & \pmark&      &      & \pmark&       \\ 
Olmedo et al.~\cite{olmedo}                         & $N$ PRs           & PR groups + sequence    & \yes &      &      &      & \yes &       \\
PRioritize~\cite{prioritize}                        & $N$ queued PRs    & merge-queue rank        & \yes &      &      &      & \yes &       \\
\midrule
\textbf{BulkPR-Bench (ours)}                        & $N$ candidate PRs & safe subset + order     & \yes & \yes & \yes & \yes & \yes & \yes  \\
\bottomrule
\end{tabular}

\par\vspace{4pt}
\parbox{\textwidth}{\footnotesize\raggedright \textbf{Columns:}
Multi=multiple distinct requested or independently authored changes within one
evaluated instance, not merely a multi-file solution to one request;
Exec=repository execution contributes to the task oracle or evaluation;
Comp=correctness or safety can depend on the accumulated composition of
multiple changes; Subset=the system makes a scored final accept/reject choice
over candidates; Order=the system chooses or ranks an integration order rather
than receiving a fixed chronology; Roll-$K$=candidates arrive in bounded
batches while earlier unresolved candidates may remain actionable.
\yes~=yes; \pmark~=coverage in only some task variants or a restricted analogue.}
\end{table*}

\textbf{Repository change and review.}
SWE-bench and its extensions evaluate one issue-level repository change with
held-out verification~\cite{swebench,multiswebench,swebenchlive}.
More recent benchmarks cover feature development, original long-horizon
objectives, complete issue-resolution cycles, and dialogue-driven
workflows~\cite{featurebench,deepswe,swemarathon,swecycle,
dialogueswebench}, broadening task duration, lifecycle coverage, and
interaction, but each instance still targets one requested outcome or issue. SWE-STEPS and ChainSWE instead evaluate fixed change
chains: repository state accumulates, but the agent neither chooses chain
membership nor reorders it~\cite{beyondisolated,chainswe}. Code-review
benchmarks likewise assess comments, decisions, or revision utility for one
PR at a time~\cite{tufano2021review,codereviewer,crabench,swereview}.
Together, these tasks establish repository reasoning and executable
verification without making the choice among concurrently pending changes
the output.

\textbf{Composition and queue management.}
Merge research studies conflict detection, resolution, and structure-aware
integration~\cite{mens2002survey,apel2011semistructured,
brun2011proactive,deepmerge}. GitGoodBench evaluates Git operations such as
conflict resolution and history rewriting, whereas MOSAIC-Bench evaluates a
fixed three-stage composition~\cite{gitgood,mosaic}. Closer to queue
management, Olmedo et al.\ group and sequence PRs from textual conflicts,
while PRioritize ranks waiting PRs to improve pipeline
throughput~\cite{olmedo,prioritize}. Among the representative designs in
Table~\ref{tab:benchmark-comparison}, none jointly evaluates an agent-chosen
final subset and integration order against composition-sensitive repository
truth under bounded rolling visibility.

\textbf{Why construction is needed.}
Historical issues, PRs, and review records expose combinations that
maintainers attempted, not counterfactual outcomes for subsets and orders
they never tried~\cite{yu2015wait,mcintosh2014review}. They therefore cannot
by themselves provide the composition-specific relation labels or exact
oracle plan required by this evaluation.
Section~\ref{sec:construction} describes how we construct and execute these
counterfactual states.

\section{The Bulk-PR Governance Task}\label{sec:task}
\subsection{Task Contract}\label{sec:task-contract}

An instance contains a frozen repository snapshot, $N$ candidate PR diffs
under neutral identifiers, public tests, and a fixed interaction budget. The
agent is asked to deliver a large safe subset in executable order. It acts
through batch decisions; the scored delivery output is the \emph{realized
merge plan} accepted by the evaluator, not a patch or a self-reported final
plan. Each decision also updates a persistent, revisable typed relation ledger
used for separate diagnostics. The ledger distinguishes pairwise and higher-order
conflicts, dependencies, all-or-none groups, must-reject candidates,
duplicates, and superseding changes.

The gold relation graph (the \emph{full gold} constraints) and exact oracle
remain on the scoring side. A
realized plan is safe when every committed atomic state satisfies the full
gold constraints, and executable when its protected diffs apply in the
submitted order and each prerequisite precedes its dependent. The runtime
public gate---the released public test suite, which we also call public
CI---exposes only public constraints, so its acceptance is necessary
for committing a proposal but is not sufficient for gold safety.

\subsection{Rolling Protocol}\label{sec:rolling}

The queue follows a frozen arrival order and is released in consecutive
batches of size $K$, except possibly the last. Future batches are absent from
the agent-visible file system, while the workspace and agent session persist;
thus $K$ changes concurrent visibility rather than memory. In the buffered
protocol, at most $B$ candidates may remain pending across batch boundaries,
each for at most $T$ boundaries after arrival and only when deferral is
renewed. In the no-deferral protocol, every candidate must be merged or
permanently rejected in its arrival batch.

At batch step $s$, the available set $A_s$ contains the new batch and any
still-pending candidates. The agent submits an ordered merge list $M_s$, a
disjoint defer set $D_s$, and an updated relation ledger, with
$M_s,D_s\subseteq A_s$. Candidates in $A_s\setminus(M_s\cup D_s)$ are
permanently rejected.

The evaluator treats $M_s$ as one atomic proposal. If the public gate accepts
it, the evaluator applies the protected diffs in the submitted order to an
authoritative trunk and commits the proposal as a whole. If the gate rejects
it, none of its members is accepted and those candidates leave the queue;
separately deferred candidates remain pending. The evaluator never repairs a
failed proposal, although the agent may use public-gate feedback and public
tests in later decisions.

Delivery metrics score accepted proposals concatenated in execution order.
A scored prefix ends at an accepted atomic-proposal boundary, not after each
individual diff within the proposal. Public-green proposals that violate
hidden gold remain in the realized plan and are penalized by scoring rather
than rolled back. Relation diagnostics separately score the revisable ledger.
A malformed or missing decision advances the step as an empty action, marks
the trial invalid, and does not prevent later batches from being released.
Appendix~\ref{app:protocol} specifies execution and integrity checks.

\section{Metrics and Diagnostics}\label{sec:metrics}

The Relational Delivery Score (RDS) is the sole ranking metric for
relational delivery. Global-SGY and Exact
Completion ask whether that localized performance becomes a deployable
whole-queue plan; \emph{RDS (hidden)}, CriticalRecall, and controlled
information conditions diagnose where performance is lost.

\subsection{Relational Delivery Score}\label{sec:sgy}

RDS measures realized delivery on candidates whose correct handling depends
on PR relations, rather than total PR throughput. It scores relation groups locally
so that one failed group does not erase correct decisions elsewhere, and
equal group and repository weights prevent large components or repositories
from dominating; relation-free delivery and whole-queue deployability are
measured separately below.

For repository $r$, the full gold graph induces relation groups
$\mathcal C_r$: constraints sharing a PR form one connected component, and
an isolated must-reject candidate forms a singleton. Candidates with no
relation constraint are excluded. For trial $t$ and group $c$, let
$\mathrm{OPT}_c$ be the largest number of group members in any safe solution,
$R_{r,t,c}$ the realized accepted trace projected onto that group, and
$G_{r,t,c}$ indicate that the projection has only safe prefixes and an
executable order. Let
$V_t=\mathrm{Valid}_t\wedge\mathrm{ExecutorCompleted}_t$. The group score is
\begin{equation}
q_{r,t,c} =
\begin{cases}
0, & \neg V_t\ \text{or}\ \neg G_{r,t,c},\\
|R_{r,t,c}|/\mathrm{OPT}_c,
& V_t\wedge G_{r,t,c}\wedge\mathrm{OPT}_c>0,\\
1, & V_t\wedge G_{r,t,c}\wedge\mathrm{OPT}_c=0.
\end{cases}
\end{equation}
Thus a safe partial delivery receives fractional credit. The last case gives
full credit for correctly accepting no candidate from a zero-optimum group;
accepting any member makes that group unsafe and therefore scores zero.
A safe realized set larger than the exact optimum is flagged as an artifact
error and excluded rather than silently clipped.

Groups are equally weighted within a run. Repeats are averaged inside each
repository before repositories are macro-averaged:
\begin{equation}
\mathrm{RDS}=\frac{1}{|\mathcal R|}\sum_{r\in\mathcal R}
\frac{1}{n_r}\sum_{t=1}^{n_r}
\frac{1}{|\mathcal C_r|}\sum_{c\in\mathcal C_r}q_{r,t,c}.
\end{equation}
This order makes repositories, rather than repeated runs, the units of the
reported score.

RDS scores the system's realized decision. A proposal rejected by the public
gate is atomically rolled back and does not enter $R_{r,t,c}$; its members
leave the queue, but later proposals over a group's deferred, still-pending
members may improve that group's score. A committed
prefix that violates a hidden gold relation zeros its group but not unrelated
groups. RDS therefore measures delivery after model decisions, public-gate
feedback, and revision---not the correctness of the model's initial,
feedback-free judgment.

\subsection{Whole-Queue and Strict Certificates}\label{sec:global-sgy}

RDS omits relation-free candidates and localizes a group failure. Global
Safety-Gated Yield (Global-SGY) instead uses the complete realized
accepted set $R_t$, including relation-free candidates. Let
$\mathrm{OPT}_{N,r}$ be the maximum safe cardinality over all $N$ candidates
in repository $r$. Its whole-queue gate is
\[
\Gamma_t=\mathrm{Valid}_t\wedge\mathrm{ExecutorCompleted}_t\wedge
\mathrm{AllPrefixSafe}_t\wedge\mathrm{Executable}_t.
\]
Here validity requires protocol-conforming decisions; executor completion
requires the episode to finish normally; prefix safety checks every committed
atomic-proposal boundary against full gold; and executability checks the
realized order.
For repository $r$,
\begin{equation}
\mathrm{Global\text{-}SGY}_{r,t}
=\Gamma_t\frac{|R_t|}{\mathrm{OPT}_{N,r}}.
\end{equation}
It uses the same repeat-then-repository aggregation as RDS. Any invalid,
unfinished, unsafe, or unexecutable run scores zero; Global-SGY is a strict
companion measure of whole-queue deployability and is never combined with
RDS.

Let $P_t=\bigcup_s M_s$ be the set of all candidates proposed for merging
during the run. Writing $\mathrm{Exact}_{r,t}$ for the per-run Exact Completion
certificate,
\begin{equation}
\mathrm{Exact}_{r,t}
=\mathds{1}\!\left[\Gamma_t\wedge |R_t|=\mathrm{OPT}_{N,r}
\wedge P_t=R_t\right].
\end{equation}
The final equality requires that no proposed merge remain unrealized.
Table~\ref{tab:main-results} reports Exact Completions as raw successes/runs.

\subsection{Relation and Information Diagnostics}
\label{sec:relmetrics}\label{sec:arms}

\emph{RDS (hidden)} applies the RDS calculation only to positive-optimum
relation groups containing a constraint invisible to public CI. It therefore
measures positive delivery under hidden interactions rather than refusal
credit. One repository (langgraph) has no eligible group, so this diagnostic
covers 17 repositories.

Let $E_r^*$ be repository $r$'s normalized gold relation atoms and
$\hat E_{r,t}$ the trial's final active typed ledger; a match requires exact
relation type, membership, and direction. For a gold atom $e$, let
$\mathcal O_e$ contain the maximum-size subsets of $e$'s original relation
component that are safe under the gold constraints with $e$ removed. Its criticality weight is
\begin{equation}
w_e =
\frac{|\{S\in\mathcal O_e:S\text{ violates }e\}|}{|\mathcal O_e|}.
\end{equation}
CriticalRecall then gives greater weight to relations whose omission admits
unsafe maximum-size decisions:
\begin{equation}
\mathrm{CriticalRecall}_{r,t} =
\frac{\sum_{e\in E_r^*\cap\hat E_{r,t}}w_e}
{\sum_{e\in E_r^*}w_e}.
\end{equation}
It is undefined when the denominator is zero and otherwise uses the same
repeat-then-repository aggregation as the delivery metrics.

\textsc{Agent} uses its own ledger. In the \textsc{Gold-fed} diagnostic, the
same model receives the reliable gold subgraph induced by candidates seen so
far but still chooses every merge, deferral, and order. A clairvoyant
\textsc{Solver} receives full gold and uses a deterministic scheduler under
the same buffered $K{=}32$ protocol. Gold feeding changes the model's
information; the solver additionally changes the information horizon and
decision maker. Their RDS differences are diagnostic headroom, not model
scores or a causal decomposition. Appendix~\ref{app:metric-details} defines
two additional solver replays retained in the artifact.

\begin{figure*}[!t]
  \centering
  \includegraphics[width=\textwidth]{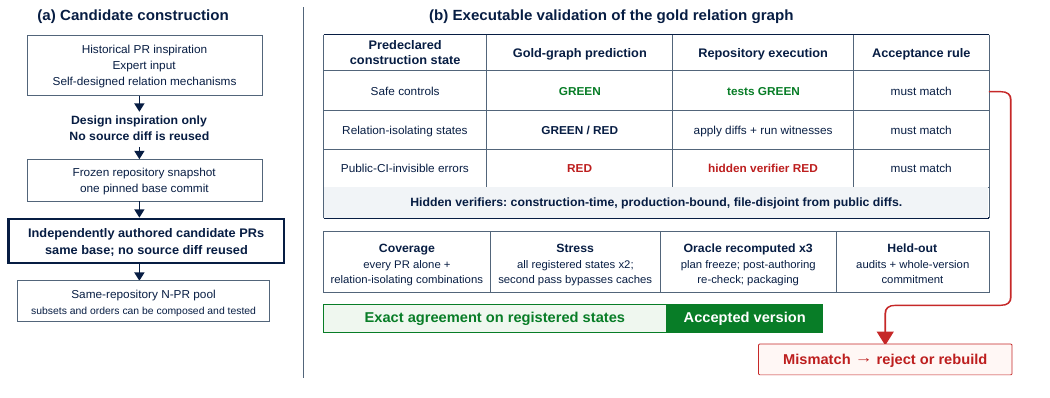}
  \Description{Two-panel construction workflow. Panel (a) uses historical
  PRs, expert input, and designed relation mechanisms as inspiration, freezes
  one repository base, independently authors candidate PRs without reusing
  source diffs, and forms a same-repository pool. Panel (b) compares
  predeclared construction-state predictions with repository execution and
  hidden verifiers.
  Coverage checks, stress runs, repeated oracle computation, and held-out
  audits accept a version only on exact state-by-state agreement; a mismatch
  causes rejection or rebuilding.}
  \caption{Executable relational construction. Historical PRs and expert input
  guide new candidate designs; candidates are independently authored on frozen
  snapshots. Registered compositions, hidden verifiers, and exact oracles must
  agree before a pool is frozen.}
  \label{fig:construction}
\end{figure*}

\section{Benchmark Construction}\label{sec:construction}

Because historical queues cannot supply counterfactual truth
(Section~\ref{sec:related}), BulkPR-Bench constructs interacting candidates
on frozen repository snapshots, executes registered compositions, validates
public-invisible relations with hidden verifiers, and freezes a pool only
when repository behavior and the proposed gold graph agree
(Figure~\ref{fig:construction}).

\subsection{Dataset, Sources, and Scope}\label{sec:dataset}

The suite contains 581 independently constructed candidate PRs in 18
repositories across Python, Go, and TypeScript/Node/Bun ecosystems, with one
pool per repository (Table~\ref{tab:dataset-summary}). The table reports $N$,
$\mathrm{OPT}_N$, relation atoms, and the fraction hidden from public tests
for every pool.

\begin{table*}[t]
\centering\normalsize
\renewcommand{\arraystretch}{1.0}
\caption{Dataset overview: 18 repository pools and 581 independently
constructed candidate PRs. Relation coverage is designed, not a natural
frequency estimate.}
\label{tab:dataset-summary}
\setlength{\tabcolsep}{3pt}
\setlength{\heavyrulewidth}{1pt}
\begin{tabular*}{0.97\textwidth}{@{\extracolsep{\fill}}
  >{\raggedright\arraybackslash}m{0.18\textwidth}
  >{\raggedright\arraybackslash}m{0.14\textwidth}
  *{7}{>{\centering\arraybackslash}m{0.078\textwidth}}
  @{}}
\toprule
Repository & Language & $N$ & OPT$_N$ & Atoms & Hidden\% & Max comp &
H-rank & Families \\
\midrule
adk-go & Go & 32 & 26 & 12 & 25\% & 4 & 3 & 5/7 \\
attrs & Python & 32 & 22 & 12 & 42\% & 3 & 2 & 3/7 \\
chi & Go & 33 & 26 & 12 & 33\% & 3 & 3 & 5/7 \\
click & Python & 32 & 23 & 12 & 42\% & 3 & 2 & 4/7 \\
eino & Go & 33 & 26 & 13 & 23\% & 4 & 2 & 4/7 \\
gin & Go & 33 & 27 & 14 & 14\% & 4 & 2 & 4/7 \\
langgraph & Python & 32 & 26 & 11 & 18\% & 3 & 3 & 5/7 \\
networkx & Python & 32 & 23 & 12 & 33\% & 3 & 2 & 3/7 \\
openai-agents-python & Python & 32 & 26 & 12 & 42\% & 3 & 2 & 4/7 \\
openclaw & TypeScript & 32 & 27 & 17 & 12\% & 7 & 2 & 4/7 \\
opencode & TypeScript & 32 & 27 & 12 & 17\% & 3 & 2 & 4/7 \\
packaging & Python & 32 & 23 & 12 & 42\% & 3 & 2 & 3/7 \\
python-sdk & Python & 33 & 26 & 12 & 25\% & 4 & 2 & 4/7 \\
rich & Python & 32 & 26 & 12 & 42\% & 3 & 2 & 4/7 \\
sqlc & Go & 33 & 26 & 13 & 23\% & 3 & 2 & 4/7 \\
vercel-ai & TypeScript & 32 & 26 & 12 & 25\% & 3 & 2 & 4/7 \\
yaml & TypeScript & 32 & 26 & 11 & 36\% & 3 & 3 & 5/7 \\
zod & TypeScript & 32 & 26 & 11 & 36\% & 3 & 3 & 5/7 \\
\bottomrule
\end{tabular*}
\par\vspace{4pt}
\parbox{\textwidth}{\footnotesize\raggedright \textbf{Columns:} $N$=pool size; $\mathrm{OPT}_N$=full-pool oracle optimum (Section~\ref{sec:global-sgy}); Atoms=declared relation atoms of the gold graph, counting singleton must-reject atoms; Hidden=fraction of atoms invisible to the public test suite; Max comp=largest connected component of the relation graph, in PRs; H-rank=largest atom arity; Families=relation families present, out of the seven in the ledger taxonomy (pairwise conflict, higher-order conflict, dependency, all-or-none group, must-reject, duplicate, supersedes).}
\end{table*}

These are \emph{constructed PRs on real repository snapshots}, not historical
open PRs: historical PRs and expert input guide task design, but no source
diff is reused. Every candidate is newly authored against the same pinned base
and must apply cleanly there, so subsets and orders can be composed without
inheriting an authoring sequence (Figure~\ref{fig:construction}(a)).
Repositories are selected for active engineering, reproducible build and test
paths, and licensing evidence rather than popularity alone.

A pool of 595 historical reference PRs provides construction inspiration and
a surface comparison, not a matched sample. Taking the median PR size within
each repository and then the median across all 18 repositories, historical
references change 3 files and 49.3 diff LOC, compared with 2 files and 20.5
diff LOC for constructed candidates (Appendix~\ref{app:release}). The
relation-family mix is deliberately designed to exercise queue interactions,
so neither these surface summaries nor the suite estimates natural relation
prevalence, semantic difficulty, or review-process
representativeness~\cite{baltes2022sampling,nagappan2013diversity}.

\subsection{Executable Ground Truth}\label{sec:truth-gates}

The gold graph is accepted only when symbolic prediction and
repository behavior agree. A \emph{state} applies an ordered subset of
candidate diffs to the snapshot. For every registered state, a constraint
predictor gives the expected green or red outcome, while the
repository-execution gate applies the diffs and runs the frozen test scope.
The registered evidence covers every nonempty subset within each relation
component, together with whole-pool witnesses and selected cross-component
probes; it does not enumerate all $2^N$ global subsets. Any mismatch rejects
the pool. Apply and infrastructure failures remain separate outcomes rather
than substitutes for an expected red state.

For each public-invisible engineered relation, a construction-time
\emph{hidden verifier} is bound to concrete production behavior. It passes on
the registered base and negative-control states and fails on the registered
target states. The accepted public projection becomes the runtime public
gate, while formal trials are scored against the frozen full-gold constraints
rather than by rerunning hidden verifiers or the full repository test scope
after every proposal. Finally, an exact solver derives
$\mathrm{OPT}_N$---exact with respect to the frozen gold constraints---and a
maximum-safe witness set; a separate topology check confirms that the witness
admits an executable order. Appendix~\ref{app:construction-audit} specifies
the repeated-execution, failure-witness, flakiness, oracle-recomputation, and
drift checks.

\subsection{Held-Out Release, Licensing, and Ethics}\label{sec:heldout}

Before formal evaluation, candidate diffs, ground truth, and hidden verifiers
remain non-public. During an episode, candidate diffs are revealed only at
their scheduled batch; future diffs, full gold, and hidden verifiers remain
outside the agent-visible tree. A committed public randomness beacon fixes neutral
identifiers and arrival order. A frozen pool version is admitted to formal
evaluation only after its execution gates pass; every evaluation is bound to
that version, and any change to version-bound bytes creates a new version
and reruns the affected leakage and integrity checks. Because
publication changes direct contamination risk, later evaluations must account
for the dated release~\cite{jacovi2023plaintext,sainz2023contamination,
livecodebench,swebenchlive}.

We will release the formal pools, hidden verifiers, scorer, and protocol as
versioned artifacts at \artifacturl. BulkPR-Bench's code will be released
under the MIT License and its constructed annotations, manifests, and task
diffs under CC~BY~4.0; upstream snapshots and upstream-derived patch context
remain governed by their original repository licenses.
The release will bind the reported results to a tag and manifest digest and
provide metadata and instructions for reproducing the reported tables from
frozen trial records. Appendix
Table~\ref{tab:appendix-repository-sources} identifies each fixed upstream
snapshot and its original repository license; the artifact manifest retains
the full commits and license-file URLs. The release omits contributor names
and emails, historical PR diffs, and expert source material; the surface audit
retains public repository/PR IDs. BulkPR-Bench evaluates
repository states rather than developers and is intended for offline agent
evaluation, not unattended production merging; its scores do not support
claims about developer populations. Evaluations run in isolated containers
without scoring-side artifacts or credentials. Repository cards record the
benchmark's motivation, composition, provenance, construction, intended use,
maintenance, and known limits~\cite{gebru2021datasheets,
bender2018datastatements,pushkarna2022datacards}.

\section{Experimental Setup}\label{sec:setup}

\subsection{Primary Profile and Sweeps}\label{sec:matrix}

The primary profile covers all 18 repositories and uses the frozen default
arrival order, generic instructions, typed ledger, buffered protocol
($B{=}4$, $T{=}16$; Section~\ref{sec:rolling}), and $K{=}32$. Each model--repository pair has three separately initialized runs
($n{=}3$). The evaluated models are Claude-Opus-4.8, DeepSeek-V4-Pro, GLM-5.2,
GPT-5.4, Kimi-K3, and Qwen3.7-Max.

All model backends use the same frozen Claude Code v2.1.138 agent scaffold,
runtime, 150-turn interaction budget, generic instructions, and ledger schema.
Because agent scaffolding can materially affect repository-task
performance~\cite{sweagent,agentless,aiagentsmatter}, the primary comparison
holds these choices fixed. Alternative prompts, ledger formats, arrival
orders, and gold-information conditions are evaluated separately and do not
enter the primary ranking. Trials hit by infrastructure failures are rerun;
turn-budget exhaustion remains a model failure and scores zero.

\begin{figure}[tbp]
  \centering
  \includegraphics[width=\columnwidth]{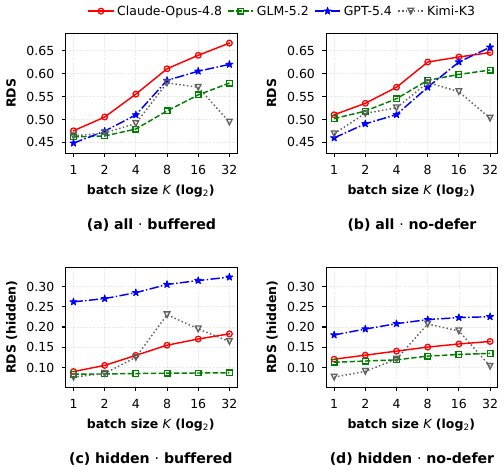}
  \Description{The figure contains four line charts of RDS against batch
  size $K$ for Claude, GLM, GPT, and Kimi. The top row covers all relation
  groups and the bottom
  row covers hidden groups; the left column uses the buffered protocol and
  the right uses no-deferral. Claude and GPT generally improve with greater
  visibility, while Kimi peaks at an intermediate $K$ and then declines;
  hidden-group trends vary by model and protocol.}
  \caption{RDS versus batch size $K$. Points average three runs per eligible
  repository (18 for RDS; 17 for RDS (hidden)). Columns compare buffered and
  no-deferral; rows use different scales. The four highest-RDS models are
  shown; Table~\ref{tab:main-results} reports all six at buffered $K{=}32$.}
  \label{fig:kcurves}
\end{figure}

RQ2 varies $K\in\{1,2,4,8,16,32\}$ under buffered and no-deferral while
holding the default order, generic instructions, and typed-ledger interface
fixed. Figure~\ref{fig:kcurves} reports three-run means at every point; its
buffered $K{=}32$ values match the primary estimates. The $K{<}32$ points
are aggregate three-run means (all six models in the artifact) and are
reported without clustered intervals. Five pools contain $N{=}33$ candidates, so their
common $K{=}32$ point does not provide full-pool visibility; $K{=}32$ is the
maximum-information operating point, and the $K$ sweep supplies the
rolling-visibility axis. The gold-fed
diagnostic uses one buffered $K{=}32$ run per model--repository pair.
Appendix~\ref{app:oracle-grid} reports the offline buffer sensitivity grid.

\subsection{Baselines and Diagnostic Arms}\label{sec:arms-view}

We include three deterministic public-information baselines:
\textsc{Merge-Queue} scans once and rejects public-gate failures;
\textsc{CI-Fixedpoint} revisits rejections after trunk changes; and
\textsc{Batch-Greedy} tries candidates within each batch without deferral.
\textsc{Merge-All}, which proposes each released batch as one atomic
proposal, and \textsc{No-Op}, which merges nothing, are anchors;
\textsc{Gold-fed} and \textsc{Solver} conditions remain unranked diagnostics
(Section~\ref{sec:relmetrics}). Each follows the frozen arrival order,
yields one deterministic outcome per repository, and sees no hidden
verifiers.

\subsection{Aggregation, Statistics, and Scope of Inference}\label{sec:stats}

Repositories are the units of aggregation and resampling. We average three
runs within each repository, then macro-average the 18 repositories equally;
PRs, batches, and repeats within one repository share a base, test
ecosystem, and construction style and are not treated as independent
repository samples.
The 95\% bias-corrected and accelerated (BCa) intervals resample these 18
means 10,000 times and capture
cross-repository, not run-to-run, variation
~\cite{efron1987bca,field2007clustered}. Baselines are fixed-suite point
estimates. Because the purposive suite is not a probability sample, its
intervals describe these pools rather than a repository population.

\section{Results}\label{sec:results}
We first compare final delivery at the primary buffered point, then examine
batch-size and deferral sensitivity, relation recovery and action, and strict
completion.

\begin{table*}[t]
\centering
\caption{Buffered is the primary protocol. $K{=}32$ results (18 repositories $\times$ 3 runs/model) are sorted by RDS; brackets are repository-level 95\% bootstrap intervals. No-deferral is the RQ2 sensitivity in Figure~\ref{fig:kcurves}.}
\label{tab:main-results}
{\normalsize
\renewcommand{\arraystretch}{1.0}
\setlength{\tabcolsep}{8pt}
\setlength{\heavyrulewidth}{1pt}
\ifdefined\bprssco\else\newlength{\bprssco}\newlength{\bprstmp}\fi
\setlength{\bprssco}{0pt}
\settowidth{\bprstmp}{RDS}\ifdim\bprstmp>\bprssco\setlength{\bprssco}{\bprstmp}\fi
\settowidth{\bprstmp}{Global-SGY}\ifdim\bprstmp>\bprssco\setlength{\bprssco}{\bprstmp}\fi
\settowidth{\bprstmp}{RDS\,(hidden)}\ifdim\bprstmp>\bprssco\setlength{\bprssco}{\bprstmp}\fi
\settowidth{\bprstmp}{CriticalRecall}\ifdim\bprstmp>\bprssco\setlength{\bprssco}{\bprstmp}\fi
\settowidth{\bprstmp}{[88.8, 88.8]}\ifdim\bprstmp>\bprssco\setlength{\bprssco}{\bprstmp}\fi
\begin{tabular}{@{}l|>{\centering\arraybackslash}p{\bprssco}>{\centering\arraybackslash}p{\bprssco}>{\centering\arraybackslash}p{\bprssco}>{\centering\arraybackslash}p{\bprssco}|c@{}}
\toprule
 & \multicolumn{4}{c|}{Repository-weighted scores (\%)} & \multicolumn{1}{c}{Certificate} \\
\midrule
Method & RDS & Global-SGY & RDS\,(hidden) & CriticalRecall & Exact Completion \\
\midrule
\multirow{2}{*}{Claude-Opus-4.8} & \textbf{66.6} & \underline{5.6} & \underline{18.3} & 57.7 & \multirow{2}{*}{\textbf{3/54}} \\[-1pt]
 & [60.4, 73.2] & [0.0, 16.7] & [9.8, 29.4] & [49.8, 65.1] &  \\[3pt]
\multirow{2}{*}{GPT-5.4} & \underline{62.0} & \textbf{13.0} & \textbf{32.4} & 52.1 & \multirow{2}{*}{\underline{2/54}} \\[-1pt]
 & [54.0, 70.2] & [4.3, 31.5] & [20.6, 47.4] & [44.7, 59.2] &  \\[3pt]
\multirow{2}{*}{GLM-5.2} & 57.9 & 2.0 & 8.7 & 54.3 & \multirow{2}{*}{1/54} \\[-1pt]
 & [51.3, 64.8] & [0.0, 7.7] & [3.4, 19.3] & [46.8, 61.2] &  \\[3pt]
\multirow{2}{*}{Kimi-K3} & 49.4 & 5.3 & 16.4 & 50.3 & \multirow{2}{*}{\underline{2/54}} \\[-1pt]
 & [41.9, 57.4] & [1.5, 12.4] & [9.2, 23.8] & [41.9, 58.0] &  \\[3pt]
\multirow{2}{*}{Qwen3.7-Max} & 47.2 & 0.1 & 6.3 & 42.9 & \multirow{2}{*}{0/54} \\[-1pt]
 & [37.2, 56.9] & [0.0, 0.4] & [2.3, 14.1] & [35.4, 50.7] &  \\[3pt]
\multirow{2}{*}{DeepSeek-V4-Pro} & 40.9 & 0.3 & 2.3 & 35.2 & \multirow{2}{*}{0/54} \\[-1pt]
 & [34.6, 47.2] & [0.1, 0.6] & [0.0, 6.2] & [29.4, 40.7] &  \\[3pt]
\midrule
\textsc{Merge-Queue} & 47.8 & 0.0 & 5.4 & --- & 0/18 \\
\textsc{CI-Fixedpoint} & 53.1 & 0.0 & 2.9 & --- & 0/18 \\
\textsc{Batch-Greedy} & 47.8 & 0.0 & 5.4 & --- & 0/18 \\
\textsc{Merge-All} & 19.0 & 0.8 & 0.0 & --- & 0/18 \\
\textsc{No-Op} & 19.0 & 0.0 & 0.0 & --- & 0/18 \\
\bottomrule
\end{tabular}
}
\par\vspace{4pt}
\begin{minipage}{\textwidth}
\footnotesize\textit{Notes.} RDS scores relation groups; Global-SGY scores the whole queue, including free PRs; RDS (hidden) covers 17 repositories. CriticalRecall is an importance-weighted, exact-typed diagnostic and is not defined for baselines. \emph{Exact Completion} reports successes/runs (54/model; 18/baseline). Bold/underline mark the top two model values in the delivery and certificate columns and do not denote significance. \textsc{No-Op} is the score obtained by merging nothing; Appendix~\ref{app:rds-composition} explains the equal baseline values. Baselines are deterministic point estimates. Both protocols remain in the machine-readable artifact.
\end{minipage}
\end{table*}

\begin{figure}[tbp]
  \centering
  \includegraphics[width=\columnwidth]{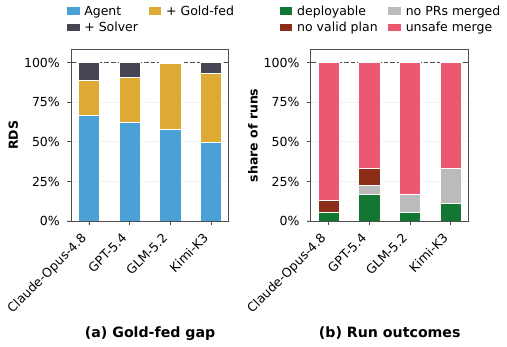}
  \Description{The figure contains two stacked bar charts of buffered
  $K{=}32$ results for four models. Panel (a) shows agent RDS, the numerical
  gap to gold-fed RDS, and the remaining gap to a full-gold buffered solver.
  Panel (b) partitions runs into deployable, no valid plan, merged nothing,
  and unsafe merge outcomes; unsafe merge is the largest category for every
  model. The stacked RDS gaps are diagnostic comparisons, not a causal
  decomposition.}
  \caption{Primary buffered $K{=}32$ diagnostics for the four highest-RDS
  models. (a) Agent RDS (three runs/cell), its gap to gold-fed RDS (one
  run/cell), and the remaining gap to the full-gold solver; the gaps are
  diagnostic, not causal. (b) Mutually exclusive outcomes (54 runs/model);
  Section~\ref{sec:rq4} aggregates all six models, whose scores are in
  Table~\ref{tab:main-results}.}
  \label{fig:recovery-action}
\end{figure}

\subsection{RQ1: Final Delivery}\label{sec:rq1}

\textbf{Three of six models scored higher than every sequential baseline in
RDS, but whole-queue delivery remained low.}
Claude-Opus-4.8, GPT-5.4, and GLM-5.2 scored 66.6\%, 62.0\%, and 57.9\% RDS
versus 53.1\% for \textsc{CI-Fixedpoint}; the other models ranged from 40.9\%
to 49.4\%. GPT-5.4 led
Global-SGY at 13.0\%, while all three sequential baselines scored zero.

\subsection{RQ2: Visibility and Batch Size}\label{sec:rq2}

\textbf{Higher-$K$ conditions generally produced higher observed mean RDS,
but effects varied by model and deferral policy.}
Under buffering, $K{=}32$ exceeded $K{=}1$ for all six models by 0.1--19.1
points: Claude, DeepSeek, GPT, and GLM gained at least 11.7 points, Kimi-K3
peaked at $K{=}8$, and Qwen was nearly flat. Hidden-group curves were
likewise heterogeneous.
At $K{=}32$, removing deferral helped GPT, GLM, and Kimi but
hurt Claude, DeepSeek, and Qwen, with changes from $+3.7$ to $-16.6$ points.
Under buffering, DeepSeek ended at 40.9\% and Qwen at 47.2\%; both remained
below all three sequential baselines at $K{=}32$.

\subsection{RQ3: Relation Recovery and Action}\label{sec:rq3}

\textbf{Agents recovered only part of the consequential relation graph, but
usually respected what they recorded.}
Buffered CriticalRecall ranged from 35.2\% to 57.7\% across all six models
(Table~\ref{tab:main-results}). Across 12 model--protocol cells, agents
respected 98.1\%--100.0\% of the relations they explicitly recovered and that
were evaluable (Appendix~\ref{app:relation-recovery-action}); this rate is
conditional on recovered relations.

Across all six models, gold-fed RDS spans 77.5\%--99.1\%, 22.3--50.2 points
above the corresponding agent RDS;
Figure~\ref{fig:recovery-action}(a) shows the four highest-RDS models. The
gap is diagnostic, not a causal recovery estimate
(Section~\ref{sec:relmetrics}). The full-gold solver also changes the
information horizon and policy, so its residual gap is not pure action loss.

\subsection{RQ4: Strict Completion and Outcomes}\label{sec:rq4}

\textbf{Exact completion was rare, and unsafe merging dominated.}
Only 8 of 324 runs met Exact Completion; none completed all three repeats for
any repository (Table~\ref{tab:main-results}).
Figure~\ref{fig:recovery-action}(b) shows the outcomes.
Across all six models, 226 runs ended with an unsafe merge (69.8\%), 59 merged nothing
(18.2\%), 27 produced deployable nonempty plans (8.3\%), and 12 had no valid
plan (3.7\%).

These four outcomes form a coarse, mutually exclusive partition that groups
the eight buckets of Appendix~\ref{app:outcomes}. \emph{Deployable} requires
a nonempty safe delivery and includes the 8 Exact Completions; \emph{merged
nothing} passes all gates without delivery; and \emph{no valid plan} covers
invalid, incomplete, or whole-queue-gate failures, none of which contained
an unsafe commit in our runs.

\section{Discussion and Limitations}\label{sec:discussion}
\textbf{The evidence does not isolate one bottleneck.}
Incomplete CriticalRecall and higher gold-fed RDS implicate missing relation
information but do not separate it from subset selection, ordering,
feedback-driven revision, or refusal. Batch size and deferral jointly change
visibility, timing, feedback, and revisability, so their curves show
configuration sensitivity rather than a causal visibility effect.

\textbf{Claims are bounded by the suite, objective, and setup.}
The 581 candidates form one designed pool on each of 18 snapshots; neither
their relation mix nor the purposive intervals represent natural queue
frequencies, other projects, or within-repository variation. The cardinality
oracle and equally weighted RDS omit business value, urgency, review cost, and
harm severity. Passing means satisfying pinned tests, hidden verifiers, and
the registered full gold, not production correctness. All backends share one
scaffold, prompt, ledger, arrival order, and turn budget
(Section~\ref{sec:matrix}), so other designs may change
results~\cite{sweagent,agentless,aiagentsmatter}.
Three runs per cell incompletely characterize run variation.

\textbf{Future work should vary objectives and agent design.}
Alternative scaffolds and a dedicated bulk-PR harness could maintain a richer
interaction graph, plan within buffer limits, and replan after gate feedback;
value-aware objectives could weight PR importance and risk.

\section{Conclusion}\label{sec:conclusion}
BulkPR-Bench evaluates queue-level PR governance---recovering consequential
relations, choosing a large safe subset, and producing an executable order
under rolling release---using execution-validated relation truth, exact
optima, and realized-plan scoring across 581 candidates on 18 repository
snapshots. Three of six models scored above the strongest
sequential baseline in RDS, yet only 8 of 324 runs achieved Exact Completion,
separating relation-group progress from dependable whole-queue governance.

\bibliographystyle{ACM-Reference-Format}
\bibliography{references}

\appendix

\section{Protocol and Metric Details}
\subsection{Execution and Integrity}\label{app:protocol}
Each released task pins the base commit, snapshot archive hash,
candidate-diff hashes, neutral identifiers, arrival order, and truth
fingerprint. Protected diffs, future candidates, signed rolling state,
and full gold remain outside the agent-visible tree. Accepted groups
execute on an authoritative trunk writable only by the root-side
evaluator, which applies the protected diffs in the submitted order
and synchronizes the resulting files back to the workspace. It does
not trust agent-controlled Git metadata, references, hooks, or claimed
merge status. Integrity failures are separated from model protocol
failures.

\subsection{Solver Replays}\label{app:metric-details}
Two solver replays use agent-recovered relations. The rolling replay
uses the active ledger after updates and retractions at each batch and
obeys the original partition, $K$, and buffering regime. The static
replay instead plans once at $K{=}N$ from the final active ledger.
A missing or incomplete timeline makes the rolling replay unavailable;
the final ledger is not used to reconstruct earlier knowledge. The
static replay is a ceiling under a different information and execution
regime, not a counterfactual trajectory of the agent.

\subsection{Construction Audit and Versioning}\label{app:construction-audit}
Registered real-gate states run twice, with caches bypassed on the
second pass; critical states receive a flakiness probe and a full-suite
confirmation. Green states must show that the intended tests executed.
Red states require an executed witness failure or a preregistered
failure signature; apply and infrastructure failures never substitute
for red. The maximum-safe-subset oracle is independently recomputed at
plan freeze, after authoring, and during task compilation. Each
accepted version pins the snapshot, diffs, truth fingerprint, test
scope, and gate transcripts. Post-freeze drift or any content change
invalidates the old task bytes and requires a new version.

\subsection{Terminal Outcome Taxonomy}\label{app:outcomes}
Every scored non-infrastructure trial enters one of eight mutually
exclusive buckets, in this priority order: turn-budget exhaustion;
invalid output or protocol failure; unsafe-light; unsafe-heavy; exact
completion; safe-all-reject; safe-near-optimal; and safe-suboptimal.
Unsafe-light violates one gold relation atom, whereas unsafe-heavy
violates at least two. Among safe, non-exact runs, safe-all-reject has
$|R|{=}0$, safe-near-optimal has $|R|{=}\mathrm{OPT}_N-1$, and
safe-suboptimal contains the remaining positive-yield outcomes. The
coarse partition of Section~\ref{sec:rq4} groups these buckets---unsafe
merge covers unsafe-light and unsafe-heavy---and additionally routes
safe runs that fail the whole-queue executability gate into \emph{no
valid plan}. These are machine-judged terminal facts, not automatic
explanations of why a run failed.

\section{Additional Diagnostics}\label{app:additional-diagnostics}

This appendix reports formal relation diagnostics, structural sensitivity,
and the composition of RDS.

\subsection{Relation Recovery and Action}
\label{app:relation-recovery-action}
\begin{figure}[H]
  \centering
  \includegraphics[width=\columnwidth]{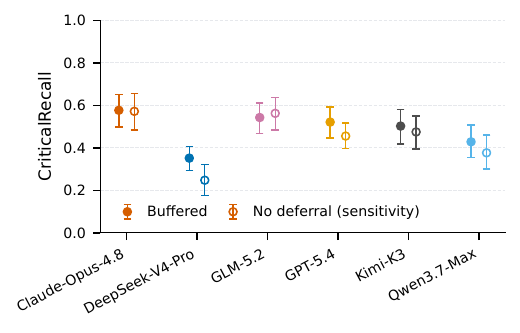}
  \Description{A dot-and-error-bar plot compares CriticalRecall at K equals
  32 for Claude-Opus-4.8, DeepSeek-V4-Pro, GLM-5.2, GPT-5.4, Kimi-K3, and
  Qwen3.7-Max. Filled markers show buffered estimates and hollow markers show
  no-deferral estimates. The buffered/no-deferral estimates are 57.7/57.2,
  35.2/24.8, 54.3/56.3, 52.1/45.6, 50.3/47.5, and 42.9/37.7 percent,
  respectively. Error bars show repository-clustered 95 percent bootstrap
  intervals.}
  \caption{\textbf{Critical-relation recovery remains incomplete.}
  CriticalRecall at $K{=}32$ under buffered (filled) and no-deferral (hollow)
  protocols. Points are repository-macro means over 18 repositories and three
  runs per model--repository cell; bars show repository-clustered 95\%
  bootstrap intervals.}
  \label{fig:recovery-full}
\end{figure}

Figure~\ref{fig:recovery-full} reports direct recovery of consequential
relations for six models under the primary buffered protocol and the
no-deferral sensitivity. Both use $K{=}32$, default order, the generic
prompt, the typed ledger, 18 repositories, and $n{=}3$ runs per cell.
We first average repeats within each repository and then macro-average
repositories; intervals use a repository-clustered 95\% bootstrap.
Structured output need not expose every latent model belief, so the
artifact retains complete counts and timing diagnostics.

Across all 12 cells,
Recovered$\rightarrow$Respected ranges from 0.981 to 1.000,
conditional on relations that were explicitly recovered and evaluable.

\subsection{RDS Composition and Equal Baselines}
\label{app:rds-composition}
Across the 18 repositories, 33 of 172 relation groups must be rejected
outright. Under repository-macro aggregation, correct refusal of those
groups gives the No-Op baseline an RDS of 0.1905; the raw group fraction
and the macro-averaged anchor are therefore close but not identical.

Two equal baseline scores in Table~\ref{tab:main-results} are structural.
\textsc{Merge-Queue} and \textsc{Batch-Greedy} accept exactly the same
candidates: both scan the arrival order once and accept a candidate
exactly when the public gate passes on the previously accepted set plus
that candidate, differing only in proposal granularity. \textsc{Merge-All}
proposes each released batch as one atomic proposal; every full-batch
proposal fails the public gate, so no relation-group member is ever
safely realized and its RDS reproduces the No-Op refusal profile. Its few
realized acceptances come from trailing single-candidate batches of
$N{=}33$ pools and add no safe relation-group delivery, so they appear
only in its nonzero Global-SGY.

\subsection{Offline Oracle Sensitivity}\label{app:oracle-grid}
We use a zero-cost clairvoyant offline oracle grid to quantify how
much of the batching-induced ceiling loss buffering can recover, and to
size the buffer. Table~\ref{tab:offline-oracle-grid} sweeps buffer
capacity $B$ against deferral horizon $T$ (in batches): the structural
deployment ceiling climbs with both and plateaus. The
$B{=}4$/$T{=}16$ configuration used for buffered model runs was chosen
from this grid: it recovers $98\%$ of the no-buffer regret at $K{=}8$
($75\%$ at $K{=}1$ and at least $97\%$ for $K{\ge}2$), while larger
buffers add little. The grid is computed from pool structure alone,
without model runs, so this choice does not condition on model
outcomes.

\section{Dataset Surface Comparison and Sources}
\label{app:release}
\subsection{Audited Surface Comparison}
We compare 595 screened historical PRs with all 581 constructed candidates
using aggregate-safe changed-file and diff-LOC summaries. Reference PRs were
stratified by month. We exclude bot-authored PRs and PRs whose changed paths
are entirely recognized dependency manifests, documentation files, or
vendor, third-party, or minified files. When a 12-month window yielded fewer
than 30 qualified PRs, it was extended in 12-month increments.
Table~\ref{tab:appendix-representativeness} first takes the median within each
repository and then the median across all 18 repositories, giving each
repository equal weight. These surface checks do not establish semantic,
difficulty, natural-frequency, or review-process representativeness.

\begin{table}[t]
\centering
\caption{\textbf{Constructed candidates are smaller on repository-median surface measures.}}
\label{tab:appendix-representativeness}
\small
\setlength{\tabcolsep}{4pt}
\begin{tabular}{@{}lcc@{}}
\toprule
Dimension & Reference & Constructed \\
\midrule
\multicolumn{3}{@{}l}{\textit{18 repositories (595/581)}} \\
Median files changed & 3 & 2 \\
Median diff LOC & 49.3 & 20.5 \\
\bottomrule
\end{tabular}
\par\vspace{3pt}
\parbox{\columnwidth}{\footnotesize Values take the median within each repository, then the median across repositories; each repository therefore receives equal weight.}
\end{table}

\subsection{Original Repository Snapshots}
Table~\ref{tab:appendix-repository-sources} gives abbreviated base
commits and upstream licenses; the artifact provides full commits and
pinned license-file links. Upstream licenses do not govern the
benchmark's own code or constructed data, whose licenses are stated in
Section~\ref{sec:heldout}.

\raggedbottom
\begin{table}[H]
\centering
\caption{\textbf{All 18 upstream snapshots are pinned.} Full commits and license URLs are in the artifact.}
\label{tab:appendix-repository-sources}
\small
\setlength{\tabcolsep}{2pt}
\begin{tabular}{@{}lll@{}}
\toprule
Repository & Base commit & Original repository license \\
\midrule
\texttt{adk-go} & \texttt{e25f1a31167f} & Apache-2.0 \\
\texttt{attrs} & \texttt{d544dd2dacb4} & MIT \\
\texttt{chi} & \texttt{8b258c7bb28f} & MIT \\
\texttt{click} & \texttt{b67832c2167e} & BSD-3-Clause \\
\texttt{eino} & \texttt{922b6a8a233b} & Apache-2.0 \\
\texttt{gin} & \texttt{34dac209ffb6} & MIT \\
\texttt{langgraph} & \texttt{49ae27c2ae98} & MIT \\
\texttt{networkx} & \texttt{b1f956344071} & BSD-3-Clause \\
\texttt{openai-agents-python} & \texttt{e29b023b5b3e} & MIT \\
\texttt{openclaw} & \texttt{41691a82d5ba} & MIT \\
\texttt{opencode} & \texttt{05c3e40a4e64} & MIT \\
\texttt{packaging} & \texttt{9d0ec47b36a1} & Apache-2.0 OR BSD-2-Clause \\
\texttt{python-sdk} & \texttt{3a6f2996cdd8} & MIT \\
\texttt{rich} & \texttt{9d8f9a372cc5} & MIT \\
\texttt{sqlc} & \texttt{e209d865845b} & MIT \\
\texttt{vercel-ai} & \texttt{6976682b5718} & Apache-2.0 \\
\texttt{yaml} & \texttt{bf03c0cb1ee9} & ISC \\
\texttt{zod} & \texttt{912f0f51b0ce} & MIT \\
\bottomrule
\end{tabular}
\end{table}

\begin{table}[H]
\centering\small
\caption{\textbf{The adopted buffer nearly reaches the offline ceiling at
$K{=}8$.} $B$ is buffer capacity and $T$ is the deferral horizon in batches.
Each entry is the equal-weight mean over 18 repositories of the
condition-specific oracle optimum divided by the full-pool optimum; bold marks the adopted
$B{=}4,T{=}16$.}
\label{tab:offline-oracle-grid}
\begin{tabular}{@{}lrrrr@{}}
\toprule
 & $T{=}1$ & $T{=}2$ & $T{=}4$ & $T{=}16$ \\
\midrule
no deferral & \multicolumn{4}{c}{0.882} \\
$B{=}1$ & 0.938 & 0.951 & 0.960 & 0.960 \\
$B{=}2$ & 0.946 & 0.970 & 0.985 & 0.985 \\
$B{=}4$ & 0.946 & 0.976 & 0.998 & \textbf{0.998} \\
$B{=}8$ & 0.946 & 0.979 & 1.000 & 1.000 \\
\midrule
OPT$_N$ & \multicolumn{4}{c}{1.000} \\
\bottomrule
\end{tabular}
\end{table}

\end{document}